\documentclass{ws-procs961x669}            % default, citations in superscript
\usepackage{xcolor}

\begin{document}
\title{News and views regarding PSR~J1757--1854, a highly-relativistic binary pulsar}

\author{A.~D.~Cameron$^*$ and M.~Bailes}

\address{ARC Centre of Excellence for Gravitational Wave Discovery (OzGrav)\\
Centre for Astrophysics and Supercomputing, Swinburne University of Technology,\\
PO Box 218, Hawthorn VIC 3122, Australia\\
$^*$E-mail: andrewcameron@swin.edu.au}

\author{V.~Balakrishnan, D.~J.~Champion, P.~C.~C.~Freire, M.~Kramer and N.~Wex}

\address{Max-Planck-Institut f{\"u}r Radioastronomie (MPIfR)\\
Auf dem H{\"u}gel 69, D-53121 Bonn, Germany}

\author{S.~Johnston}

\address{Australia Telescope National Facility, CSIRO Space and Astronomy\\
PO Box 76, Epping NSW 1710, Australia}

\author{A.~G.~Lyne and B.~W.~Stappers}

\address{Jodrell Bank Center for Astrophysics, University of Manchester\\
Alan Turing Building, Oxford Road, Manchester M13 9PL, UK}

\author{M.~A.~McLaughlin, N.~Pol and H.~Wahl}

\address{Department of Physics and Astronomy, West Virginia University\\
PO Box 6315, Morgantown, WV 26506, USA}

\author{C.~Ng}

\address{Dunlap Institute for Astronomy \& Astrophysics, University of Toronto\\
50 St. George Street, Toronto, ON M5S 3H4, Canada}

\author{A.~Possenti}

\address{Istituto Nazionale di Astrofisica (INAF), Osservatorio Astronomico di Cagliari\\
Via della Scienza 5, 09047 Selargius (CA), Italy}

\author{A.~Ridolfi}

\address{Istituto Nazionale di Astrofisica (INAF), Osservatorio Astronomico di Cagliari\\
Via della Scienza 5, 09047 Selargius (CA), Italy}
\address{Max-Planck-Institut f{\"u}r Radioastronomie (MPIfR)\\
Auf dem H{\"u}gel 69, D-53121 Bonn, Germany}

\begin{abstract}
We provide an update on the ongoing monitoring and study of the highly-relativistic double neutron star binary system PSR~J1757--1854, a 21.5-ms pulsar in a highly eccentric, 4.4-hour orbit. The extreme nature of this pulsar’s orbit allows it to probe a parameter space largely unexplored by other relativistic binary pulsars. For example, it displays one of the highest gravitational wave (GW) luminosities of any known binary pulsar, as well as the highest rate of orbital decay due to GW damping. PSR~J1757--1854 is also notable in that it is an excellent candidate for exploring new tests of General Relativity and other gravitational theories, with possible measurements of both Lense-Thirring precession and relativistic orbital deformation (through the post-Keplerian parameter $\delta_\theta$) anticipated within the next 3--5 years.

Here we present a summary of the latest interim results from the ongoing monitoring of this pulsar as part of an international, multi-telescope campaign. This includes an update of the pulsar’s long-term timing and post-Keplerian parameters, new constraints on the pulsar’s proper motion and corresponding Shklovskii kinematic correction, and new limits on the pulsar’s geodetic precession as determined by monitoring for secular changes in the pulse profile. We also highlight prospects for future work, including an updated timeline on new relativistic tests following the introduction of MeerKAT observations.%, \review{as well as a brief discussion of the pulsar’s potential detectability within the LISA band}.
\end{abstract}


\bodymatter

\section{Introduction}

For many decades, relativistic binary pulsars have been one of the key tools for studying different theories of gravity in the strong field regime, the current paradigm being Einstein's theory of General Relativity (GR). Discovered in 2016 as part of the HTRU-S Galactic Plane pulsar survey\cite{kjvs+10,cck+18}, the 21.5-ms pulsar PSR~J1757--1854 ranks as one of the most extreme examples of this class. Its high eccentricity ($e\simeq0.606$) and compact orbit (measured via the projected semi-major axis, $x\simeq2.24\,\text{lt-s}$) around a neutron star companion combine some of the best properties of other notable relativistic binary pulsars, including the high eccentricity of PSR~B1913+16\cite{ht75} (the 7.75-hr orbital period \textit{`Hulse-Taylor pulsar'}, with $e\simeq0.617$) and the compact orbit of PSR~J0737--3039A/B\cite{bdp+03,lbk+04} (the 2.45-hr orbital period\textit{`Double Pulsar'}, with $x\simeq1.42\,\text{lt-s}$). The extreme nature of PSR~J1757--1854 is further demonstrated by multiple records set by the pulsar, including the highest acceleration ($\sim700\,\text{m\,s}^{-2}$) and highest relative velocity ($\sim1000\,\text{km\,s}^{-1}$) seen in a binary pulsar, as well as the highest rate of orbital decay due to gravitational wave damping and one of the shortest predicted merger times of any Galactic double neutron star system ($\sim76\,\text{Myr}$). Together, these properties make PSR~J1757--1854 a highly promising candidate for providing new insights into theories of gravity.

Here, we describe some of the ongoing aspects of the work involving this pulsar, in advance of a more detailed peer-reviewed publication anticipated by mid-2022.

\section{Current timing and post-Keplerian tests}
Since its discovery in 2016, PSR~J1757--1854 has been observed by multiple radio telescopes as part of an international collaboration, including the Parkes\footnote{Also known by its indigenous Wiradjuri name \textit{`Murriyang'}.} (Australia), Effelsberg (Germany), Lovell (UK), Green Bank (GBT; USA) and MeerKAT (South Africa) telescopes. This data was analysed using a standard pulsar timing technique\cite{lk05}, wherein summed, average pulse profiles are cross-correlated against high signal-to-noise (S/N) template profiles to produce a dataset of high-precision pulse `times of arrival' (TOAs). Each TOA represents the mean arrival time of an integrated pulse profile from the pulsar. These are then iteratively fit using a \textit{timing ephemeris} in order to develop a high-precision model of the pulsar's behaviour.

Currently, the GBT accounts for the overwhelming majority of PSR~J1757--1854's TOAs, having produced over 22,000 TOAs since it began observing in mid-2016. Data from the GBT is recorded in a full-Stokes coherent search mode at both L and S-bands (800 MHz bandwidth centered at 1500 and 2000\,MHz respectively). We currently record one 4.4-hr orbit at each frequency every two months.

\begin{table}
\tbl{Latest timing parameters for PSR~J1757--1854, based on data from the GBT and Parkes between MJD 57405--59363 and employing the DDH\cite{fw10} binary model. Values in parentheses give the 1-$\sigma$ uncertainties on the final digit.}
{\begin{tabular}{ll}
\toprule
\multicolumn{2}{l}{\textbf{Astrometric \& spin parameters}}\\
Right ascension, $\alpha$ (J2000)\dotfill & 17:57:03.78412(2) \\
Declination, $\delta$ (J2000)\dotfill & $-$18:54:03.359(4) \\
Spin period, $P$ (ms)\dotfill & 21.4972318900292(6)\\
Spin period derivative, $\dot{P}$\dotfill & $2.627335(9)\times10^{-18}$\\
Dispersion measure (pc $\text{cm}^{-3}$)\dotfill & 378.203(2) \\
Proper motion in RA, $\mu_\alpha$ (mas $\text{yr}^{-1}$)\dotfill & $-$4.36(12)\\
Proper motion in DEC, $\mu_\delta$ (mas $\text{yr}^{-1}$)\dotfill & $-$0.8(14)\\
Period and position epoch (MJD) & 57701 \\
\\
\multicolumn{2}{l}{\textbf{Orbital parameters}}\\
Orbital period, $P_\text{b}$ (d)\dotfill & 0.183537835854(7) \\
Eccentricity, $e$\dotfill & 0.6058171(3) \\
Projected semimajor axis, $x$ (lt-s)\dotfill & 2.2378060(15) \\
Epoch of periastron, $T_{0}$ (MJD)\dotfill & 57700.92599421(3)\\
Longitude of periastron, $\omega$ ($^\circ$)\dotfill & 279.34090(9)\\
\\
\multicolumn{2}{l}{\textbf{Post-Keplerian parameters}}\\
Rate of periastron advance, $\dot{\omega}$ ($^\circ\,\text{yr}^{-1}$)\dotfill & 10.36498(3) \\
Einstein delay, $\gamma$ (ms)\dotfill & 3.5891(16) \\
Orbital period derivative, $\dot{P}_\text{b}$\dotfill & $-$5.294(6)$\times10^{-12}$ \\
Orthometric amplitude, $h_3$ ($\mu\text{s}$)\dotfill & 5.08(18) \\
Orthometric ratio, $\varsigma$\dotfill & 0.899(10) \\
\\
\multicolumn{2}{l}{\textbf{Mass measurements (based on $\dot{\omega}$ and $\gamma$)}}\\
Pulsar mass, $m_\text{p}$ ($\text{M}_\odot$)\dotfill & 1.3406(5)\\
Companion mass, $m_\text{c}$ ($\text{M}_\odot$)\dotfill & 1.3922(5)\\
Inclination angle, $i$ ($^\circ$)\dotfill & 85.0(2)\\
\botrule
\end{tabular}}
\label{tab: timing parameters}
\end{table}

\begin{figure}
\begin{center}
\includegraphics[width=0.7\textwidth]{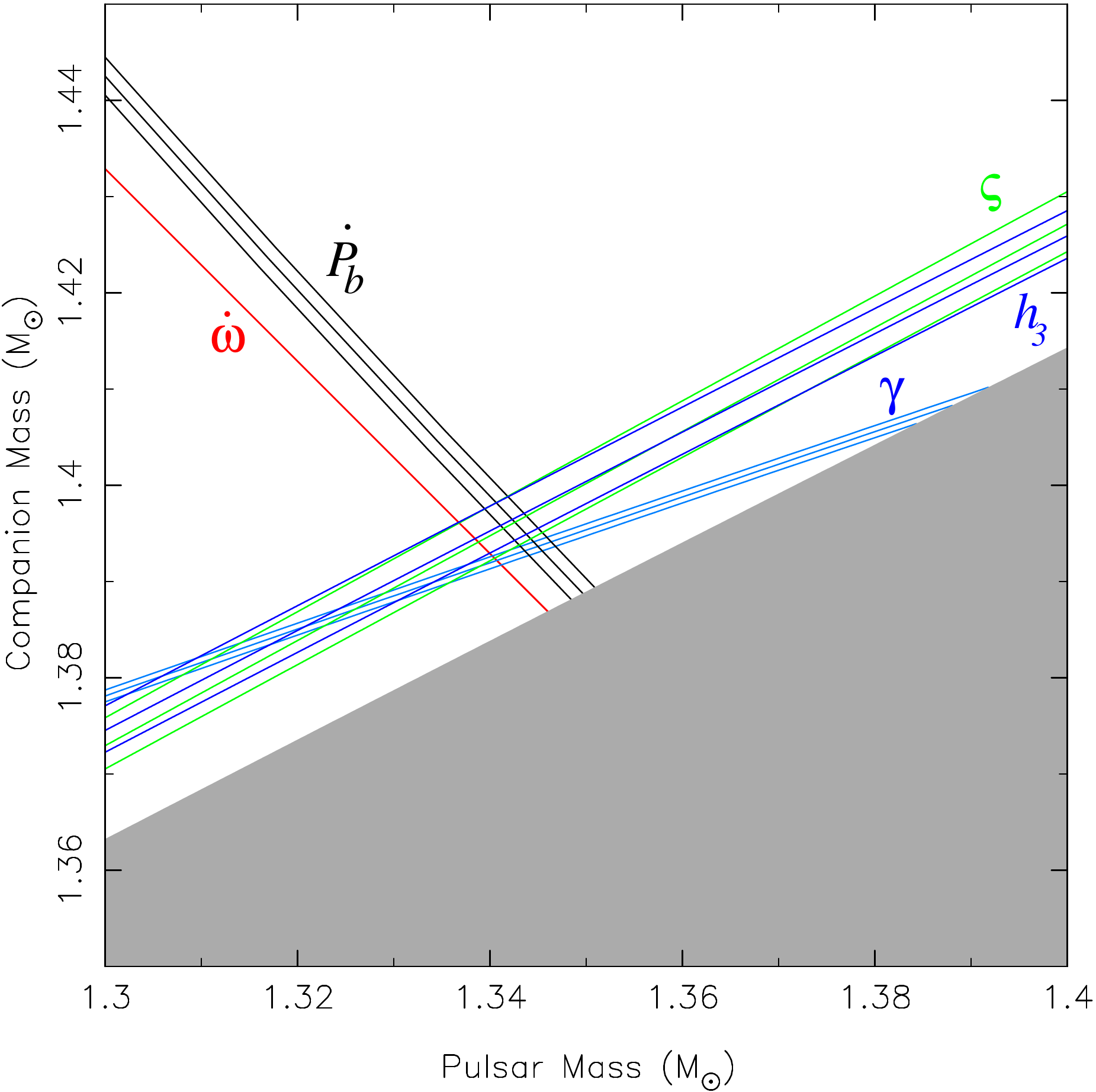}
\caption{Mass-mass diagram for PSR~J1757--1854. Each coloured triplet of lines shows the constraints (with 1-$\sigma$ uncertainty) placed by each of the measured PK parameters in Table~\ref{tab: timing parameters} on the mass of the pulsar and the companion neutron star, in this case according to GR. The measured uncertainty of $\dot{\omega}$ is so small that it cannot be seen at this scale. The grey region in the bottom right is excluded due to orbital geometry.}
\label{fig: mass-mass}
\end{center}
\end{figure}

The most recent timing ephemeris for PSR~J1757--1854 is given in Table~\ref{tab: timing parameters}. This includes five theory-independent post-Keplerian (PK) relativistic parameters, from which we can derive three independent tests of gravity. A mass-mass diagram showing the mass constraints imposed by each PK parameter under GR is shown in Figure~\ref{fig: mass-mass}. By fixing a solution for the masses of the pulsar and the companion neutron star using the intersection of $\dot{\omega}$ and $\gamma$ (the most precisely measured PK parameters), the further intersection of each remaining PK parameter provides one additional test of consistency. Both the orthometric amplitude and ratio parameters of the Shapiro delay\cite{fw10} ($h_{3}$ and $\varsigma$) are consistent to within 3.7\% and 1.9\% of their GR-predicted values respectively, and are approximately consistent with the $\dot{\omega}-\gamma$ intersection. However, although the observed value of the orbital period derivative $\dot{P}_\text{b}$ is also consistent with GR to within at least 1\%, its location on the mass-mass diagram is inconsistent with the constraints set by $\dot{\omega}$ and $\gamma$ under the assumption of GR being correct. This is further discussed in Section~\ref{subsec: pbdot}.

\subsection{Proper motion and the radiative test of gravity}\label{subsec: pbdot}
A recent development in the timing of PSR~J1757--1854 has been the first constrained measurement of the pulsar's proper motion, which is also listed in Table~\ref{tab: timing parameters}. The component in right ascension ($\mu_\alpha$) is detectable at approximately 36\,$\sigma$, while the component in declination ($\mu_\delta$) remains poorly constrained, although its contribution to the total proper motion remains small given its value relative to $\mu_\alpha$. Together, these values indicate a total proper motion of
\begin{equation}\label{eqn: proper motion}
    \mu_\text{T} = \sqrt{\mu_\alpha^{2} + \mu_\delta^{2}} = 4.4\left(3\right)\,\text{mas\,yr}^{-1}.
\end{equation}
%(**AP**){\it I suggest to expand the error bar to the total proper motion via propagation of the errors, otherwise the significance of the total proper motion is not crystal clear}  (*AP*)

This allows us to attempt to quantify the contributions contaminating the observed value of the orbital period derivative, $\dot{P}_\text{b,obs}$. We define the excess contribution as\begin{equation}\label{eqn: pbdot excess}
\dot{P}_\text{b,exs} = \dot{P}_\text{b,obs} - \left(\dot{P}_\text{b,GR} + \dot{P}_\text{b,Gal} + \dot{P}_\text{b,Shk}\right),
\end{equation}where $\dot{P}_\text{b,GR}$ is the expected intrinsic contribution from GR (here calculated using the mass values derived from the observed $\dot{\omega}$ and $\gamma$ as given in Table~\ref{tab: timing parameters}); $\dot{P}_\text{b,Gal}$ is the contribution from the acceleration of the pulsar within the Galactic potential\cite{dt91} (dependent on position, distance and the chosen model of the Galactic potential); and $\dot{P}_\text{b,Shk}$ is the contribution from the Shklovskii effect\cite{shklovskii70} (dependent on distance and proper motion).

\begin{figure}
\begin{center}
\includegraphics[width=0.85\textwidth]{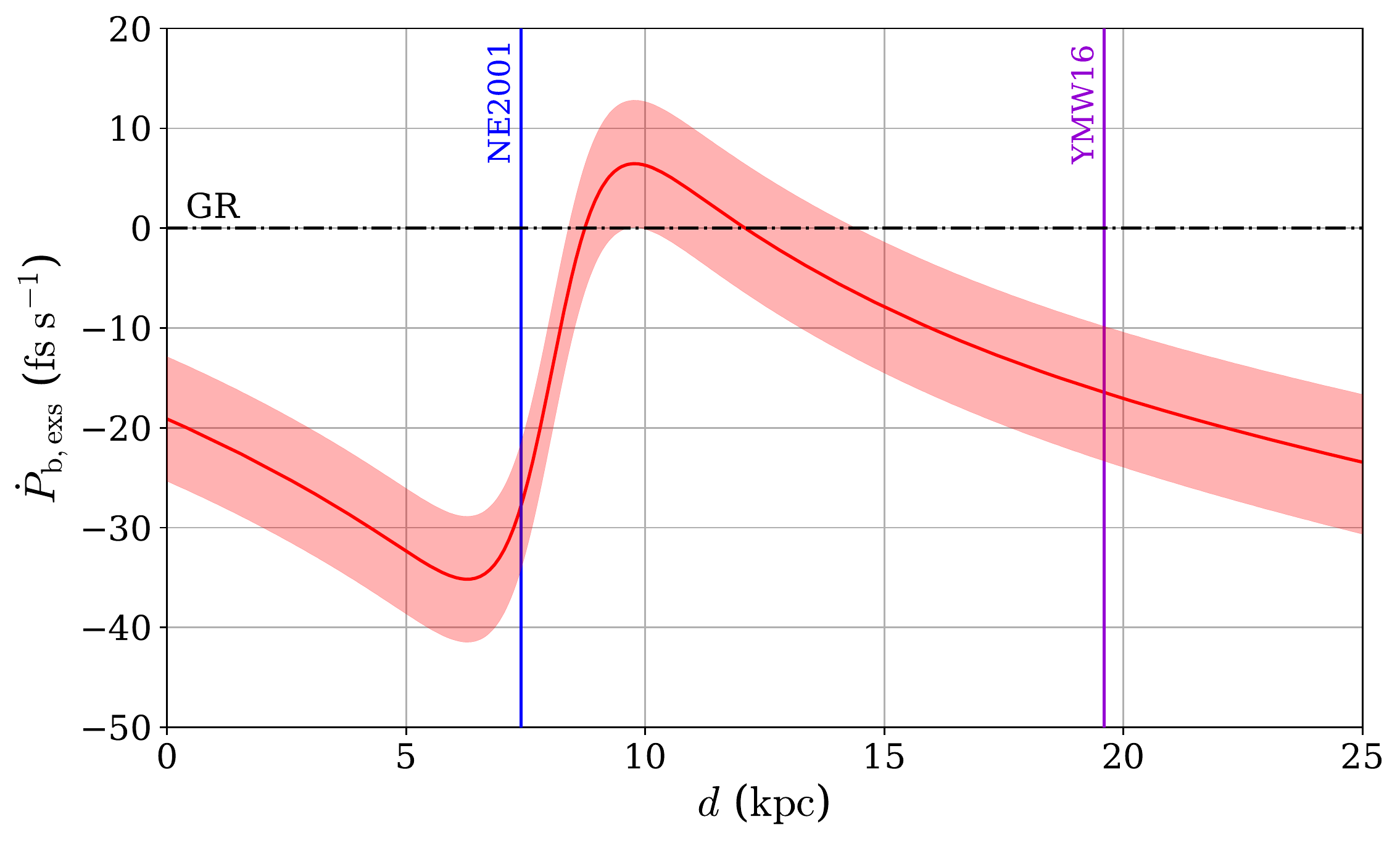}
\caption{The excess contribution to the orbital period derivative $\dot{P}_\text{b,exs}$ as calculated by Equation~\ref{eqn: pbdot excess}, plotted (red) as a function of distance. The shaded region shows the 1-$\sigma$ uncertainty. $\dot{P}_\text{b,Gal}$ was calculated using the McMillan Galactic potential\cite{mcmillan17}. The vertical blue (left) and magenta (right) lines indicate the NE2001\cite{NE2001a} and YWM16\cite{YMW16} DM-distance estimates respectively. The horizontal dashed black line shows the GR expectation of $\dot{P}_\text{b,exs} = 0$.}
\label{fig: pbdot excess}
\end{center}
\end{figure}

For the radiative $\dot{P}_\text{b}$ test of GR to be considered `passed', $\dot{P}_\text{b,exs}$ must be consistent with zero, given that the other contributions have been accurately accounted for. However, although we have now quantified the pulsar's proper motion, we still lack an accurate estimate of the pulsar's distance. This is demonstrated in Figure~\ref{fig: pbdot excess}, where we show how $\dot{P}_\text{b,exs}$ depends on the distance to PSR~J1757--1854. Assuming GR, the distance to PSR~J1757--1854 should fall between approximately 8.5--14.5\,kpc. Therefore, available distance estimates based on the pulsar's dispersion measure (DM), including both the NE2001\cite{NE2001a} (7.4\,kpc) and the YMW16\cite{YMW16} (19.6\,kpc) electron density models, are currently excluded. This is also shown in Figure~\ref{fig: pbdot excess}. However, we note that these DM-distance estimates typically come with large uncertainties\cite{ccb+20}, making a definitive determination difficult. Additionally, the McMillan Galactic potential model\cite{mcmillan17} we have used in our analysis has known shortcomings at distances of approximately 7--9\,kpc in the central regions of the Galaxy near PSR~J1757--1854's position, further complicating this assessment.

Without an accurate distance to PSR~J1757--1854, the utility of the radiative test of gravity is unfortunately limited. As seen in Figure~\ref{fig: pbdot excess}, the unknown distance introduces an approximate uncertainty to $\dot{P}_\text{b,exs}$ of at least $\pm20\times10^{-15}$, more than three times the current uncertainty of $\dot{P}_\text{b,obs}=6\times10^{-15}$, such that the precision of the radiative test of GR is limited to about 0.4\,\%.  We are currently exploring ways of further refining the estimated distance to PSR~J1757--1854 (e.g. through VLBI astrometry). However, by assuming GR, we will still be able to inform our understanding of the pulsar's evolution. For example, the currently allowed distance range indicates a transverse velocity based on $\mu_\text{T}$ between 180--300\,km\,s$^{-1}$, well in agreement with simulations of the neutron star kick associated with the second progenitor supernova\cite{cck+18}.

\section{Attempts to detect the presence of geodetic precession}

Our current understanding of the binary evolution of PSR~J1757--1854 suggests the strong likelihood of a misalignment between the pulsar's spin vector and the orbital angular momentum vector\cite{cck+18}. As a result, we anticipate the presence of geodetic precession at a rate of $\Omega_\text{geod}\simeq3.07\,^{\circ}\,\text{deg}\,\text{yr}^{-1}$, which if detected may serve as another test of GR. We have therefore analysed the current GBT dataset for evidence of changes in the pulse profile, which would indicate a changing line-of-sight through the emission cone of the pulsar as expected from geodetic precession\cite{bailes88}.

Our method adapts an approach previously applied\cite{ppb+21} to PSR~J1022+1001. Each GBT observation (at both L and S-bands) spanning greater than 80\% the orbital period is summed in time, frequency and polarisation before being fit by a template composed of three component Gaussian curves. We then normally re-sample each profile bin using the off-pulse root-mean-squared noise value to generate at least 1,000 randomly `re-noised' profiles, to which the template is re-fit. From each new fitted template, the positions and amplitudes of the major and minor peaks are then measured, as well as the intercept positions at 10\% and 50\% of the profile's maximum amplitude. The mean and standard deviation of each parameter's statistical distribution across all fitted templates are then taken as the nominal value and its uncertainty.

\begin{figure}
\begin{center}
\includegraphics[width=\textwidth]{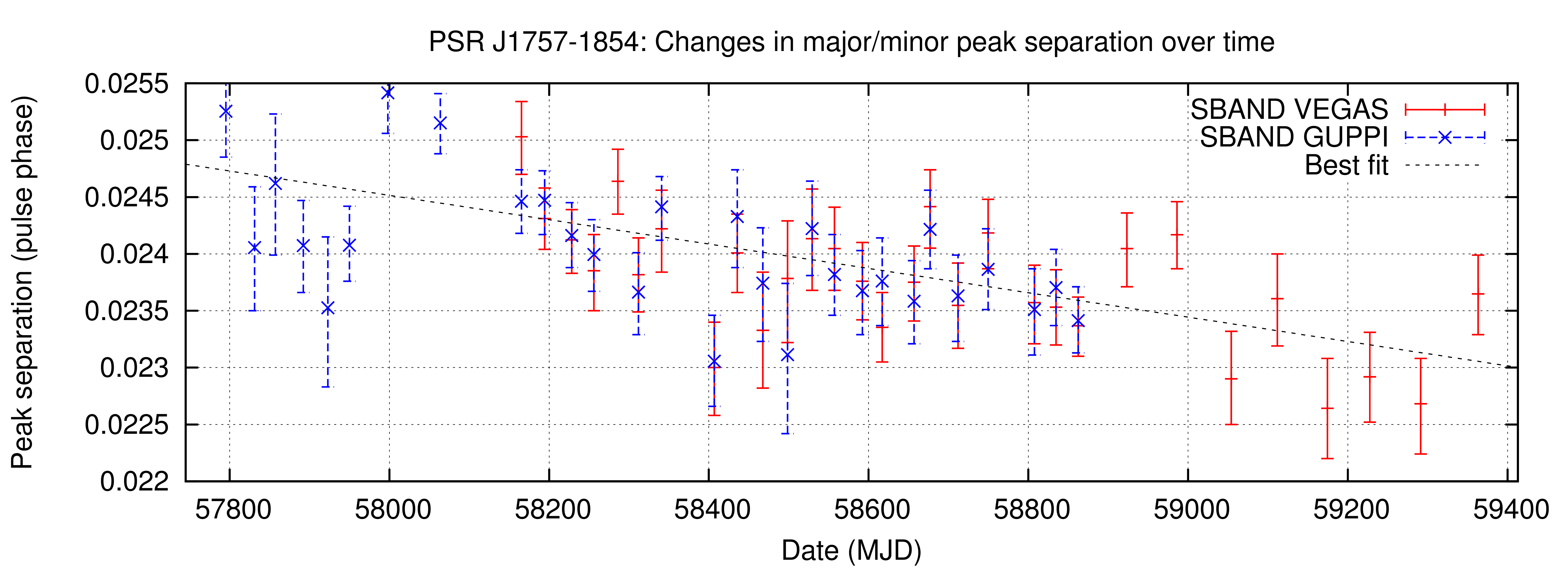}
\caption{Change in phase separation of the major and minor peaks of PSR~J1757--1854's pulse profile, as measured by the GBT at S-band using the GUPPI and VEGAS backends. The line of best fit corresponds to a decreasing separation of $0.14\left(2\right)\,^\circ\,\text{yr}^{-1}$.}
\label{fig: profile change}
\end{center}
\end{figure}

This approach has for the first time confirmed the presence of statistically significant profile change. An example of this is given in Figure~\ref{fig: profile change}, which shows that the separation between the two peaks of PSR~J1757--1854's profile (termed the `major' and `minor' peaks) is decreasing (i.e. moving closer together) by an average rate of approximately $0.14\left(2\right)\,^\circ\,\text{yr}^{-1}$. The width of the profile at both 10\% and 50\% also appears to be shrinking at a commensurate rate, with all trends most pronounced at S-Band due to the decreased influence of scattering (with respect to lower-frequency L-band observations). The precise physical implications of these changes are to be explored in a future publication, but their detection provides fundamental evidence for the presence of geodetic precession, as well as evidence for the suspected spin/orbit misalignment present in the binary system.

\section{Searches for the companion neutron star}

%A positive detection of the neutron star companion to PSR~J1757--1854 as an active radio pulsar would immensely enhance the scientific utility of this pulsar.

If PSR~J1757--1854's neutron star companion was confirmed to be an active radio pulsar, it would immensely enhance the scientific utility of this pulsar and its binary system. This is most obviously demonstrated by the prior example of the Double Pulsar, where the presence of the second pulsar PSR~J0737--3039B allowed additional tests of gravity beyond those normally available, including the detection of geodetic precession\cite{bkk+08} and an independent determination of the mass ratio\cite{ksm+06}. %Furthermore, although PSR~J0737--3039B has precessed out of our line-of-sight since 2008, the interaction of its magnetosphere with the emission of PSR~J0737--3039A and the additional constraints this can place on tests of gravity continues to be the subject of intense study\cite{ksv+21}.

We have therefore conducted an ongoing search for evidence of pulsations from the neutron star companion to PSR~J1757--1854. This search has focused on observations from the GBT, which are recorded in a coherently-dedispersed search mode so as to allow for optimal search sensitivity across all pulse periods. Given our precise knowledge of PSR~J1757--1854's orbit as detailed in Table~\ref{tab: timing parameters}, each observation is re-sampled using the inferred orbit of the companion via the custom software package \textsc{pysolator}\cite{pysolator}. This allows the companion to appear stationary such that any pulsations would appear as a single narrow peak in a resulting Fourier spectrum. To account for the small degree of uncertainty in the companion orbit, different re-sampled trials are produced using values of the mass ratio $q=m_\text{p}/m_\text{c}$ between $0.9554 < q < 0.9661$ in increments of $0.00005$. The resulting Fourier power spectra from each observation are then stacked and summed, boosting the total S/N of any pulsations.

To date, this search has unfortunately returned negative results, although the complete GBT dataset has yet to be analysed. In total, approximately 389.7\,hr of GBT observations are currently on file, although there is significant overlap in this dataset between the VEGAS and GUPPI backends during intervals when both were recorded simultaneously. Ongoing searches for the companion neutron star will continue, as the companion may become detectable in the event that it precesses into view in the near future, given the expected precession rate of $\Omega_\text{geod}\simeq2.97\,^{\circ}\,\text{deg}\,\text{yr}^{-1}$.

\section{Future observations and anticipated tests of gravity}

\subsection{Addition of MeerKAT}

Although observations with South Africa's MeerKAT telescope have been ongoing since 2019 as part of the MeerTIME program\cite{bja+20}, all these observations took place with MeerKAT's L-band receiver, which has a large 856\,MHz bandwidth centered at 1283.5\,MHz. This frequency band is located at lower frequencies than those of the GBT L-band receiver and therefore is much more susceptible to pulse broadening via scattering than the GBT L-band\cite{ksv+21}. As a consequence, even though MeerKAT can achieve about twice the S/N of the GBT for this target, the available timing precision from both telescopes is roughly equivalent.

These effects are expected to be mitigated as part of the ongoing rollout of a fleet of MeerKAT S-band receivers (with a bandwidth of approximately 1750\,MHz centred at 2625\,MHz), designed by the MPIfR. The decreased influence of scattering at these higher frequencies combined with the wider bandwidth and the pulsar's relatively flat spectral index suggest an improvement in TOA precision to within 10\,$\mu\text{s}$ or better (compared with a typical TOA precision of 25--30\,$\mu\text{s}$ currently achievable using similar TOA integration times with the current GBT and MeerKAT receivers). This will provide a significant advantage in the ongoing timing of the pulsar.

\subsection{Future tests of gravity}

PSR~J1757--1854 is expected to provide new insight into two rarely-explored tests of gravity within the next few years. The first of these is the \textit{relativistic orbital deformation}, characterised by the PK parameter $\delta_\theta$, which describes the deviation of the orbit from a pure Keplerian ellipse due to the effects of relativity. The measurability of $\delta_\theta$ scales with both eccentricity and $\dot{\omega}$, positioning PSR~J1757--1854 as an ideal system for providing constraint in the short-term. Meanwhile, the Hulse-Taylor pulsar\cite{wh16} (lower $\dot{\omega}$) and the Double Pulsar (lower eccentricity) have to date only weakly constrained $\delta_\theta$ despite having been studied for far longer. 

The second anticipated test involves the detection of Lense-Thirring precession, which to date has not been detected in a double neutron star binary system. We intend to detect this effect via the measurement of a change in the semi-major axis\cite{dt92}, $\dot{x}_\text{LT}$. Although in general $\dot{x}_\text{LT}$ becomes negligible for small misalignment angles between the pulsar spin and orbital angular momentum vectors, our recent detection of geodetic precession confirms that a significant misalignment is likely. The Lense-Thirring effect is also proportional to the neutron star moment of inertia; should the Lense-Thirring effect be detected here, it would lead to a rare measurement of this quantity and thereby provide insight into the neutron star equation of state.

The timescale over which each of these parameters is anticipated to be constrained was determined by a simulation of the anticipated timing campaigns. A simulation was produced by extrapolating the current GBT timing campaign until the end of 2024, and including an equivalent MeerKAT S-band campaign comprising one full orbit recorded every two months, starting in 2022. Timing ephemerides were fit to the data at set intervals, while averaging the results of 30 realisations of the dataset. Based on these simulations, we anticipate the achievement of a 3-$\sigma$ constraint of $\delta_\theta$ between approximately 2024--2025, and of $\dot{x}_\text{LT}$ between 2025--2026.

\section*{Acknowledgments}

The Parkes radio telescope is part of the Australia Telescope National Facility which is funded by the Australian Government for operation as a National Facility managed by CSIRO. We acknowledge the Wiradjuri people as the traditional owners of the Observatory site. The Green Bank Observatory is a facility of the National Science Foundation operated under cooperative agreement by Associated Universities, Inc. The MeerKAT telescope is operated by the South African Radio Astronomy Observatory, which is a facility of the National Research Foundation, an agency of the Department of Science and Innovation. Pulsar research at the Jodrell Bank Centre for Astrophysics and the observations using the Lovell Telescope are supported by a consolidated grant from the STFC in the UK. This work is also based on observations with the 100-m telescope of the Max-Planck-Institut f{\"u}r Radioastronomie at Effelsberg. AC and MB acknowledge OzGrav ARC grant CE170100004. AP and AR acknowledge that (part of) this work has been funded using resources from the research grant “iPeska” (P.I. Andrea Possenti) funded under the INAF national call Prin-SKA/CTA approved with the Presidential Decree 70/2016. AP and AR also acknowledge the support from the Ministero degli Affari Esteri e della Cooperazione Internazionale - Direzione Generale per la Promozione del Sistema Paese - Progetto di Grande Rilevanza ZA18GR02. AR acknowledges continuing valuable support from the Max-Planck Society. MAM and HW are members of the NANOGrav Physics Frontiers Center and supported by NSF awards 1430284 and 2020265.

\bibliographystyle{ws-procs961x669}
\bibliography{ws-procs961x669}

\end{document}